\documentclass[aps,prb,twocolumn,showpacs,floatfix,footinbib]{revtex4}
\usepackage{graphicx,graphics}
\usepackage{dcolumn}
\usepackage{amsmath,amssymb,amsfonts,latexsym,verbatim}
\usepackage{bm,bbm}
\usepackage{color}
\def\be{\begin{equation}}
\def\ee{\end{equation}}
\def\nn{\nonumber}
\def\ber{\begin{eqnarray}}
\def\eer{\end{eqnarray}}
\def\rv{{\bf r}}

\def\jv{{\bf j}}
\def\xv{{\bf {\rm x}}}
\def\Bv{{\bf B}}
\def\Av{{\bf A}}

\def\kv{{\bf k}}

\def\sigmav{\mbox{\boldmath $\sigma$}}

\def\nubold{\mbox{\boldmath $\nu$}}
\def\Nubold{\mbox{\boldmath ${\bar{\cal V}}$}}
\def\Fbold{\mbox{\boldmath ${\cal B}$}}
\def\Bbold{\mbox{\boldmath ${\bar {\cal B}}$}}
\def\Abold{\mbox{\boldmath ${\cal A}$}}
\def\Jbold{\mbox{\boldmath ${\cal J}$}}

\def\nablabold{\mbox{\boldmath $\nabla$}}
\begin{document}

\title{Gauge-Invariant Formulation of Spin-Current-Density Functional Theory}

\author{Saeed H. Abedinpour}
\email{abedinpour@gmail.com}
\altaffiliation{Present address: School of Physics, Institute for Research in Fundamental Sciences (IPM), 19395-5531 Tehran, Iran}
\affiliation{Department of Physics and Astronomy, University of Missouri, Columbia, Missouri 65211, USA}
\author{G. Vignale}
\affiliation{Department of Physics and Astronomy, University of Missouri, Columbia, Missouri 65211, USA}
\author{I. V. Tokatly}
\affiliation{IKERBASQUE, Basque Foundation for Science, E-48011, Bilbao, Spain}
\affiliation{ETSF Scientific Development Centre, 
Dpto. F\'isica de Materiales, Universidad del Pa\'is Vasco, Centro de
F\'isica de Materiales CSIC-UPV/EHU-MPC, Av. Tolosa 72, E-20018 San 
Sebasti\'an, Spain}
\affiliation{Moscow Institute of Electronic Technology, Zelenograd, 124498 Russia}
\date{\today}

\begin{abstract}
Spin-currents and non-abelian gauge potentials in electronic systems can be treated by spin-current-density functional theory, whose main input is the exchange-correlation (xc) energy expressed as a functional of spin-currents.  Constructing a functional of spin currents that is invariant under U(1)$\times$SU(2) transformations is a long-standing challenge.  We solve the problem by expressing the energy as a functional of a new variable we call ``invariant vorticity".   As an illustration we construct the xc energy functional for a two-dimensional  electron gas with linear spin-orbit coupling and show that it is proportional to the fourth power of the spin current. 
\end{abstract}


\maketitle

\section{Introduction}\label{intro}
In the last few decades, density functional theory (DFT)~\cite{ref:vignale_book} has 
grown to be a widely used method for studying the ground-state properties of interacting many-electron systems and the range of its applications has been expanding.
An important generalization of DFT, which takes into account the orbital effects of an external magnetic field, is the non-relativistic {\it current density functional theory} (CDFT), which was formulated by Vignale and Rasolt  (VR)~\cite{ref:vignale87,ref:vignale88,ref:vignale90} in the late eighties. 
The basic variable of that theory  (in addition to the usual particle and spin densities) is the {\em paramagnetic current density}  $\jv_p(\rv)$, which has the following advantages upon the ``physical current density" $\jv(\rv)$: (i) it has no explicit dependence on the external vector potential -- a property it shares with previous relativistic formulations~\cite{ref:rajagopal73,ref:eschrig85}, and (ii) it does not vanish  in the limit of uniform density and magnetic field, which is vital to the construction of a local density approximation (LDA).   Unfortunately $\jv_p$ is not a gauge-invariant variable.  VR ensured the gauge invariance of the exchange-correlation (xc) energy by expressing the latter in terms of the vorticity 
\be\label{VRVorticity}
\nubold := \nablabold\times \frac{\jv_p}{n}\,,
\ee
which {\it is} gauge invariant. While this choice is not unique (any gauge-invariant field which is in one-to-one correspondence to the vorticity would be in principle admissible), it was shown to be the most natural one, in the sense of leading to a local description of the effect of the magnetic field in the quasi-homogeneous limit.

In extending their theory to spin polarized systems,\cite{ref:vignale88} VR proved the analogue of the Hohenberg-Kohn theorem for paramagnetic spin-currents, and observed that the xc energy $E_{xc}$ -- now a functional of paramagnetic spin currents -- should be invariant under local U(1) $\times$ SU(2) gauge transformations.  These are transformations in which the wave function is acted upon by the position-dependent operator  $\exp[i {\bf \Lambda}(\rv)]$,  where ${\bf \Lambda}(\rv)$ is a matrix in two-dimensional spin space.  However, the behavior of the paramagnetic  spin currents under these transformations is quite complicated and VR were unable to identify a variable  analogous to the vorticity, such that an $E_{xc}$ functional of this variable would be  manifestly  gauge invariant. 

Recently, Bencheikh~\cite{ref:bencheikh} and G\"orling~\cite{ref:rohra07} have further extended  spin-CDFT to incorporate spin-orbit (SO) interactions.  These interactions are naturally described in terms of  SU(2) gauge potentials, i.e., vector potentials that are matrices in spin space. 

In Ref.~\onlinecite{ref:tokatly08} it was noticed that in many realistic situations the SU(2) vector potential produced by the SO coupling has a nonvanishing covariant curl, which is equivalent to the presence of a non-Abelian magnetic field. This effective magnetic field generates ``diamagnetic" spin currents in the ground state of molecules and solids with non-negligible SO interaction. In spite of the fact that the ground state spin current is hard to detect experimentally,  its inclusion into the set of basic variables of DFT is important to correctly account for the corresponding contribution to the energetics.  SU(2) gauge potentials also appear in effective hamiltonians for spin-transfer torque systems~\cite{ref:tatara}, and in pseudospin-orbit coupled systems like graphene~\cite{ref:castroneto}. 

 Bencheikh derived the form of the Kohn-Sham equation in terms of effective U(1) $\times$ SU(2) gauge potentials, both expressed as functional derivatives of a gauge-invariant $E_{xc}$.  But again, no explicitly SU(2)-invariant expression for $E_{xc}$ was provided.  While the main challenge of CDFT is to construct practical approximations for the xc functional,  such as the exact-exchange current- and spin-current density functionals recently developed in Refs.~\onlinecite{ref:rohra07,ref:rohra06} and~\onlinecite{ref:helbig08}, we believe it is of fundamental importance to know how to construct manifestly gauge-invariant xc functionals of the spin-current densities.

In this paper we present a solution to this problem. 
First we note that both the external scalar $U$ and vector $\Av$ potentials and the SO terms can be combined into a U(1)$\times$SU(2) four-vector gauge potential, collectivelly denoted by $\Abold$, in such a way that the exact many-body Hamiltonian becomes invariant under local U(1)$\times$SU(2) gauge transformations. This leads to the next key observation that the ground state energy $E$  depends on $\Abold$  only through the invariant field strength  $\Bbold[\Abold]$, defined as the ``SU(2)-invariant curl" of  $\Abold$:  all these quantities will be defined precisely below.  $\Bbold$ is invariant under SU(2) rotations in spin space and transforms as a pseudo-vector in the ordinary space, so we can write 
\be\label{ExcA}
E[\Abold]=\bar{E}[\Bbold[\Abold]]\,.
\ee
where $\bar{E}$ is a scalar functional of $\Bbold$. 
Notice that in the familiar spinless case  $\Abold$ reduces to the usual vector potential $\Av$  and  $\Bbold$ reduces to the magnetic field $\Bv=\nablabold \times\Av$, so the above equation simply says that the energy is a functional of the magnetic field.  
We will show below that the quantity ${\cal N}^{-1} \Jbold_p$, where $\Jbold_p$ denotes the  paramagnetic particle/spin currents and ${\cal N}$ is the particle/spin density (again, precise definitions will be given below),  behaves under U(1) $\times$ SU(2) gauge transformations precisely like the  gauge potential  $\Abold$.    This suggests a path to constructing a gauge-invariant xc energy functional $E_{xc}$.  Let us introduce the ``SU(2)-invariant vorticity"  
\be\label{vorticity}
\Nubold:=\Bbold[{\cal N}^{-1}\Jbold_p]\,,
\ee
which is the ``SU(2)-invariant curl" of  ${\cal N}^{-1}\Jbold_p$ and will be shown to reduce to the VR vorticity~(\ref{VRVorticity}) in the spinless case.  Since ${\cal N}^{-1} \Jbold_p$ transforms as $\Abold$ it is evident that $\Nubold$ is invariant under SU(2) rotations in spin space,  just as $\Bbold[\Abold]$ is.  Therefore, the gauge invariance of $E_{xc}$ will be guaranteed if we write it as a scalar functional  of the SU(2)-invariant vorticity (hereafter abbreviated as ``invariant vorticity"):
\be \label{ExcJ}
E_{xc}[\Jbold_p] =\bar{E}_{xc}[\Nubold]\,
\ee
(the dependence on particle and spin density is left implicit -- see concluding section). 

The functional dependence of $E_{xc}$ on $\Nubold$ is of course  very complex and nonlocal.  The test of usefulness of  Eq.~(\ref{ExcJ})  is whether there are limiting cases in which the $E_{xc}$ functional  can naturally be cast in this form. We will show that this is the case.  Namely, in the important limit of electron gas subjected to {\em uniform and weak} SU(2)  potentials the xc energy can be written as a quadratic local functional of the invariant vorticity.   Remarkably, in this regime the invariant vorticity is a quadratic functional of the spin currents themselves, not of their derivatives.   This opens the way to a genuine LDA, completely different in character from the local approximation of Refs.~\onlinecite{ref:vignale87} and~\onlinecite{ref:vignale88}  in which the xc energy depended on derivatives of the paramagnetic current and/or density.  The leading contribution to the present LDA functional goes as the fourth power of the spin current, and the xc potentials are proportional to its third power.

This paper is organized as follows. In Section~\ref{basic} we introduce our model and basic variables. In Section~\ref{scdft} we present the gauge invariant spin-current-density functional theory and the self-consistent Kohn-Sham equations. Our local density approximation for $E_{xc}$ is presented in Section~\ref{lda}. In Section~\ref{rashba} and Appendix~\ref{app:rashba} we calculate the exchange-correlation energy of Rashba system, which would provide the necessary input for our LDA scheme in the case of isotropic SU(2) vector potentials. Finally, Section~\ref{cncl} summarizes our main conclusions.

\section{Basic definitions and gauge transformations}\label{basic}
We start with the many-body Pauli Hamiltonian~\cite{ref:frohlich93}, which describes non-relativistic electrons in a static electromagnetic field ($\hbar=e=m=c=1$)
\begin{align}\label{eq:pauli0}
{\hat  H}&=\int \mathrm{d}{\bf r}\, \Psi^{\dagger}\left\{\frac{1}{2}{\bf \Pi}^2+\frac{1}{8}\left[{\bf \Pi}\cdot\left(\sigmav\times {\bf E}\right)+\left(\sigmav \times {\bf E}\right)\cdot{\bf \Pi}\right]\right.\nonumber\\
&\left.-U+\frac{1}{2}{\bf B}\cdot \sigmav\right\} \Psi+{\hat W}\,,
\end{align}
where $\Psi^{\dagger}=(\psi^{\dagger}_\uparrow,\psi^{\dagger}_\downarrow)$ is the two-component field operator, ${\bf \Pi}=-i\nablabold+{\bf A}$ is the kinetic momentum, $\sigmav$ is the vector of the Pauli matrices, $U$ is the electrostatic potential, ${\bf A}$ is the vector potential,  ${\bf E}=-\nablabold U$ and ${\bf B}=\nablabold\times{\bf A}$ are the electric and the magnetic fields, respectively.
The first term on the right hand side of Eq.~(\ref{eq:pauli0}) is the kinetic energy, the second is the spin-orbit coupling energy, the third and the fourth are the potential and the Zeeman energies, respectively.  ${\hat W}$ denotes the electron-electron interaction, which we assume to be spin-independent and gauge-invariant. 

We can write the hamiltonian more compactly by introducing the
U(1)$\times$SU(2) gauge potential ${\cal A}_{\mu}={\cal A}_{\mu}^{\alpha}\tau^{\alpha}$, where $\vec \tau$ is a four-vector constructed by a $2\times 2$ unit matrix $\mathbb{I}$ and Pauli matrices: $(\mathbb{I},\sigma^x,\sigma^y,\sigma^z)$ Here and in the following upper indices refer to charge ($\alpha=0$) and spin ($\alpha=x,y,z$), and lower indices to time ($\mu=0$) and space ($\mu=x,y,z$) components. A sum over repeated indices is implied~\cite{foot:index}. The components ${\cal A}_{\mu}^{\alpha}$ of the potential are defined as follows
\begin{eqnarray}
\label{A-U(1)}
{\cal A}_{0}^{0} &=& -U - \frac{{\bf E}^2}{16}, \quad {\cal A}_{i}^{0}=A_i\,, \nonumber \\
\label{A-SU(2)}
{\cal A}_{0}^{a} &=& \frac{1}{2}B_{a}, \quad  {\cal A}_{i}^{a} = - \frac{1}{4}\varepsilon_{ija}E_j\,.
\end{eqnarray}
We also introduce the covariant space derivatives ${\cal D}_{j}=\partial_j+i{\cal A}_j^\alpha\tau^\alpha$, where $j=x,y,z$, $\partial_j$ is the derivative with respect to the corresponding variable. With this notation the  Hamiltonian~(\ref{eq:pauli0}) takes the form
\begin{equation} \label{eq:pauli}
{\hat H}=\int\mathrm{d}{\bf r} \left[{\cal A}^{\alpha}_0 {\Psi}^{\dagger} \tau^{\alpha}{ \Psi} + \frac{1}{2}({\cal D}_i { \Psi})^{\dagger} ({\cal D}_i { \Psi})  \right]+{\hat W}\,.
\end{equation}
One can easily show that this Hamiltonian is invariant under a local time-independent  U(1)$\times$ SU(2) gauge transformation~\cite{ref:frohlich93,ref:bohm}:
\begin{align}\label{eq:gauge}
\Psi &\longrightarrow {\cal U}  \,\Psi \,, \nonumber\\
{\cal A}_{\mu}& \longrightarrow {\cal U}{\cal A}_{\mu}{\cal U}^{\dagger}+i{\cal U}^{\dagger}\partial_{\mu} {\cal U}\,,
\end{align}
with ${\cal U}=\exp{\left[i \Lambda^{\alpha}({\bf r})\tau^{\alpha}\right]}$, where $\Lambda^{\alpha}({\bf r})$ is an arbitrary time independent function. This transformation corresponds to the multiplication by a local phase factor supplemented with a local rotation in the spin space. Explicitly the matrix trasformation law (\ref{eq:gauge}) translates to the following transformation for the components ${\cal A}_{\mu}^{\alpha}$ of the gauge potential
\begin{equation}
\label{gauge2}
{\cal A}_{\mu}^{\alpha} \longrightarrow{\cal R}^{\alpha\beta}({\vec \Lambda})\left\{{\cal A}_{\mu}^{\beta}+\frac{i}{2}\,{\rm Tr}\left[\tau^{\beta}{\cal U}^{\dagger}\partial_{\mu} {\cal U}\right]\right\}\,,
\end{equation}
where the trace ${\rm Tr}$ is taken over $\tau$, and ${\cal R}({\vec \Lambda})$ is a $4\times 4$ matrix, with ${\cal R}^{00}=1$, ${\cal R}^{0a}={\cal R}^{a0}=0$, and ${\cal R}^{ab}=R^{ab}({\bf \Lambda})$, $R^{ab}({\bf \Lambda})$ being the $3\times3$ rotation matrix in spin space by angle $\theta=2|{\bf \Lambda}|$ along the direction ${\bf \Lambda}$. Note that ${\cal U}^{\dagger}\partial_{0}{\cal U}=0$ for static transformations.

A crucial quantity for our purposes is the ``field strength" defined as the {\it covariant curl} of the gauge potential~\cite{foot:fieldstrength} 
\begin{equation}\label{eq:field}
{\cal B}_{i} = {\cal B}_{i}^{\alpha}\tau^\alpha= \epsilon_{ijk}{\cal D}_{j} {\cal A}_{k}^{\alpha}\tau^\alpha\,,
\end{equation} 
or, more explicitly
\be\label{CovariantCurl}
{\cal B}_{i}^{\alpha}[\Abold]=\epsilon_{ijk}\left[\partial_{j} {\cal A}_{k}^{\alpha}-(1-\delta_{\alpha,0})\epsilon_{\alpha b c}{\cal A}^b_{j} {\cal A}^c_{k}\right]\,.
\ee
The ``charge" component ($\alpha=0$) of the field strength coincides with the usual magnetic field, while its ``spin" part ($\alpha=x,y,z$) corresponds to a non-Abelian SU(2) magnetic field. Here we must emphasize  an important difference between the familiar Abelian U(1) magnetic field $\Bv$ of electrodynamics and its non-Abelian counterpart ${\cal B}$.  While $\Bv$ is gauge-invariant {\it tout-court}, our non-Abelian field strength ${\cal B}$ is {\it gauge-covariant}, i.~e., under the gauge transformation it transforms as ${\cal B}\mapsto {\cal U}{\cal B}{\cal U}^{\dagger}$. In other words the behavior of each spatial component ${\cal B}_{i}$ under SU(2) rotations is that of a vector in spin space: ${\cal B}_{i}^{\alpha}\rightarrow {\cal R}^{\alpha\beta}({\vec \Lambda}){\cal B}_{i}^{\alpha}$.  The lack of full gauge invariance becomes a problem when we attempt to construct a non-local, yet gauge-invariant energy functional, for fields at different points in space transform differently.  For example, the  plausible functional
\be\label{TrialExpression}
\int d\rv\int d\rv' K(\rv,\rv'){\cal B}^a_i(\rv){\cal B}^a_i(\rv')\,
\ee
is not SU(2)-invariant, in spite of the fact that all indices are duly contracted.

\begin{figure}
\begin{center}
\includegraphics[width=1.0\linewidth]{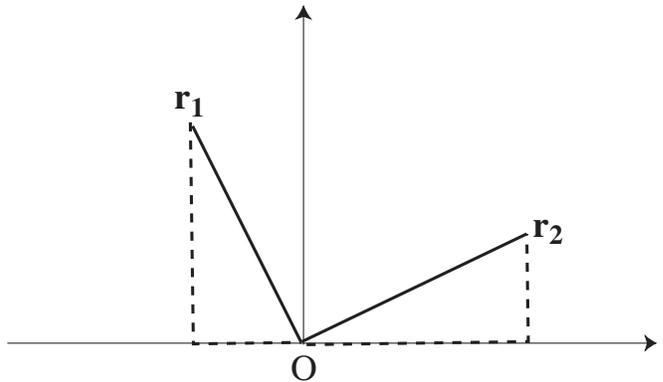}
\caption{Two admissible choices of paths connecting the origin to two points $\rv=\rv_1$ and $\rv_2$ for the evaluation of the path-ordered integral in Eq.~(\ref{deflink}).  The solid path is a straight radial line, the dashed path proceeds along the $x$-axis first, and then along the $y$-axis.} 
\label{Fig0}
\end{center}
\end{figure}

To avoid this difficulty we must go a step further and introduce an SU(2)-invariant field, which behaves like a scalar under SU(2) rotations.
To this end, we define a {\it link operator} $\hat L(\rv)$ ( a $2\times 2$ matrix) in the following manner:
\be\label{deflink}
\hat L(\rv)= P {\rm exp}\left(-i\int_{\bf 0}^{\rv}\hat A_i (\xv) dx_i\right)\,,
\ee
where the integral is taken along some conventionally specified path which connects the origin (${\bf 0}$)  to the point $\rv$.  These paths should be smooth functions of $\rv$: for example a set of lines radiating from the origin, or a set  of ``L-shaped" lines (see figure~\ref{Fig0})  would be acceptable.  

The symbol $P$ denotes the ``path ordering operator" along the chosen path that connects ${\bf 0}$ to $\rv$.  In analogy with the time-ordering operator, $P$ orders $\rv$-dependent matrices in such a way that the ones with the ``earlier" $\rv$ (i.e., closer to the origin) act before the ones with the ``later" $\rv$. Geometrically, the operator $\hat L(\rv)$ of Eq.~(\ref{deflink}) can be interpreted as a {\it parallel transport operator} -- it describes rotation of a spinor under a ``parallel transport" along the specified path in the space with nonzero SU(2) connection ${\cal A}_{i}^{a}$ (see, for example, Ref.~\onlinecite{ref:itzykson}).  It is not difficult to show that under SU(2) transformation the link operator transforms as follows:
\ber\label{linktransformations}
\hat L(\rv) &\to& \hat{\cal U}(\rv)\hat L(\rv)\nn\\
\hat L^\dagger(\rv) &\to& \hat L^\dagger(\rv) \hat{\cal U}^\dagger(\rv)\,.
\eer
Then we define the invariant field $\Bbold$ as follows:
\be\label{InvariantB}
\bar{\cal B}_i(\rv)=\hat L^\dagger(\rv) {\cal B}_i(\rv)\hat L(\rv)\,.
\ee
Making use of the previously derived transformations, it is easy to check that $\Bbold$ is indeed invariant under local SU(2) transformations.  The reason for this property is that Eq.~(\ref{InvariantB}) is a ``closed loop" form. In fact, the link operators $\hat L(\rv)$ and $\hat L^\dagger(\rv)$ in Eq.~(\ref{InvariantB}) describe a parallel transport from the point ${\bf 0}$ to $\rv$ and back, respectively. It is also worth noting that our invariant fied $\Bbold$ is closely related to the ``twisted curvature" entering the non-Abelian Stokes theorem \cite{ref:broda2001}.  The ground-state energy functional is therefore naturally expressed in terms of scalar combinations such as $\bar{\cal B}_{i}^{a}\bar{\cal B}_{i}^{a}$, which are also invariant under global spatial rotations~\cite{foot:symmetry}. In particular, the expression~(\ref{TrialExpression}) becomes SU(2)-invariant if  ${\cal B}_i^a(\rv)$ is replaced by   $\bar{\cal B}_i^a(\rv)$.

The physical spin-current densities
\begin{equation}\label{SpinCurrentDefinition}
j_{\mu}^{\alpha}({\bf r})=\left\langle\frac{\delta { {\hat H}}}{\delta {\cal A}_{\mu}^{\alpha}({\bf r})}\right\rangle = \frac{\delta E[{\cal A}]}{\delta {\cal A}_{\mu}^{\alpha}({\bf r})}\,,
\end{equation}
are SU(2)-covariant in the sense that  $j_{\mu}^{\alpha}({\bf r})\rightarrow {\cal R}^{\alpha\beta}({\vec \Lambda})j_{\mu}^{\alpha}({\bf r})$. Their $\mu=0$ components form the particle/spin density 
\begin{equation}\label{eq:spin}
j_{0}^{\alpha}({\bf r}) = s^{\alpha}({\bf r})= \left\langle\Psi^{\dagger}\tau^{\alpha} \Psi \right\rangle\, ,
\end{equation}
where $s^0({\bf r})=n({\bf r})$ is the particle density. The $\mu\neq 0$ components are particle ($\alpha=0$) and spin ($\alpha \neq 0$) currents
\begin{equation}\label{eq:current}
\begin{split}
j_{i}^{\alpha}({\bf r}) &=\frac{-i}{2}\left\langle\left[ { \Psi}^{\dagger}\tau^{\alpha}{\cal D}_{i}{ \Psi} -({\cal D}_i { \Psi})^{\dagger} \tau^{\alpha} { \Psi}\right] \right\rangle \\
&=j_{p\,i}^{\alpha}({\bf r}) + {\cal N}^{\alpha\beta}({\bf r}) {\cal A}^{\beta}_{i}({\bf r}) \, ,
\end{split}
\end{equation}
where 
${\cal N}^{\alpha\alpha}=n$, ${\cal N}^{0a}={\cal N}^{a0}=s^a$, and all the other components are zero.
The {\it paramagnetic spin-currents} $j_{p\,i}^{\alpha}$ are defined as~\cite{ref:vignale_book,ref:tokatly08}
\begin{equation}\label{eq:para_current}
j_{p\,i}^{\alpha}({\bf r},t)=-\frac{i}{2}\left\langle\left[ { \Psi}^{\dagger}\tau^{\alpha} \partial_{i}\Psi -\left(\partial_i \Psi\right)^{\dagger} \tau^{\alpha} { \Psi} \right]\right\rangle\,,
\end{equation}
and we also define $j_{{ p}\,0}^{\alpha}({\bf r}) = s^{\alpha}({\bf r})$. 
These paramagnetic currents are {\it not} gauge-covariant, but transform under gauge transformation~(\ref{eq:gauge}) in the following manner:
\begin{equation}
{j}_{p\,i}^{\alpha}\longrightarrow{\cal R}^{\alpha\beta}\left\{j_{p\,i}^{\beta}-\frac{i}{2}{\cal N}^{\beta\gamma}\,{\rm Tr}\left[\tau^{\gamma}{\cal U}^{\dagger}\partial_{i} {\cal U}\right]\right\}\,.
\end{equation}
More importantly for the present purpose,  the combination
$v^{\alpha}_{i}=({\cal N}^{-1})^{\alpha\beta}{j}^{\beta}_{p\,i}$, which is the analogue of a velocity field, transforms exactly like the gauge potentials ${\cal A}^{\alpha}_i$, namely
\begin{equation}\label{Velocity}
v_{i}^{\alpha}\longrightarrow{\cal R}^{\alpha\beta}\left\{v_{i}^{\beta}-\frac{i}{2}\,{\rm Tr}\left[\tau^{\beta}{\cal U}^{\dagger}\partial_{i} {\cal U}\right]\right\}\,.
\end{equation}
It then follows that the covariant curl of  $v_{i}$
\be\label{CovariantVorticity1}
{\cal V}_{i} = {\cal V}_{i}^{\alpha}\tau^\alpha= \epsilon_{ijk}
{\cal D}_{v,j}v_{k}^{\alpha}\tau^\alpha\,,
\ee 
or, more explicitly,
\be\label{CovariantVorticity2}
{\cal V}_i^{\alpha}=\epsilon_{ijk}\partial_{j}v_k^\alpha-(1-\delta_{\alpha,0})\epsilon_{ijk}\epsilon_{\alpha b c}v_j^b v_k^c
\ee
transforms as ${\cal V}_{i}^{\alpha}\rightarrow {\cal R}^{\alpha\beta}({\vec \Lambda}){\cal V}_{i}^{\alpha}$, i.e. it is a vector in spin space, just as the spin density.   
In Eq.~(\ref{CovariantVorticity1}) we have introduced ${\cal D}_{v,j}=\partial_j+i v_j^\alpha\tau^\alpha$.

The gauge co-variance of  ${\cal V}_{i}$ (or the spin density for that matter) poses, as noted above, a problem in constructing nonlocal functionals that are invariant under SU(2) transformation.  One way to solve the problem is to introduce,  in parallel with the introduction of $\Bbold$, the SU(2)-invariant vorticity:
\be\label{InvariantNu}
\bar{\cal V}_i(\rv)=\hat L_v^\dagger(\rv) {\cal V}_i(\rv)\hat L_v(\rv)\,,
\ee
where $\hat L_v$ is defined, in analogy to Eq.~(\ref{deflink}), as
\be\label{defvlink}
\hat L_v(\rv)= P {\rm exp}\left(-i\int_{\bf 0}^{\rv}\hat v_i (\xv) dx_i\right)\,.
\ee
In exactly the same way we can define the SU(2)-invariant spin-density
\be\label{InvariantS}
\hat{\bar s}(\rv)=\hat L_v^\dagger(\rv) \hat s(\rv)\hat L_v(\rv)\,,
\ee
where $\hat s(\rv)=s^a(\rv) \tau^a$ is the ordinary (covariant) spin density, defined in Eq.~(\ref{eq:spin}). 
Curiously, the expressions for the invariant vorticity and spin density, although invariant, contain the non-invariant paramagnetic currents through $\hat v_i (\xv)$.  We will return to this point in the conclusions. We now have all the ingredients for constructing a SU(2)-invariant xc energy functional.

\section { Spin-current-density functional theory}\label{scdft}
The ground state energy and spin current distribution of an interacting many-body system, subjected to external gauge potentials ${\cal A}_{\mu}^{\alpha}$, is obtained from the minimization with respect to $j_{p\,\mu}^{\alpha}$ of the functional
\begin{equation}\label{eq:e_gs}
E_{\rm GS}[j_{p\,\mu}^{\alpha}]=\min_{\psi\rightarrow j_{p\,\mu}^{\alpha}({\bf r})} 
\langle \psi|{ H}|\psi\rangle
=F[j_{p\,\mu}^{\alpha}]+ E_{ext}[j_{p\,\mu}^{\alpha}]\,.
\end{equation}
Here $F[j_{p\,\mu}^{\alpha}]$ is the part of the ground-state energy that does not depend on the external fields. This functional is traditionally expanded as   $F=T_s+E_{\rm H}+E_{xc}$, where $T_s$, $E_{\rm H}$ and $E_{xc}$ are the non-interacting kinetic, Hartree and exchange-correlation contributions, respectively.  $E_{ext}[j_{p\,\mu}^{\alpha}]$ , on the other hand, is the part of the ground-state energy that pertains to the coupling of the currents with the external potentials. This functional has the same form in both interacting and non-interacting systems.  
Since $F+E_{ext}$, $T_s+E_{ext}$ and $E_{\rm H}$ are separately gauge invariant, the exchange-correlation energy alone $E_{xc}=F-T_s-E_H$ has to be gauge invariant as well.  Unfortunately, the paramagnetic current variable  $j_{p\,\mu}^{\alpha}$ does not possess this invariance.  This becomes a problem when one attempts to construct approximations in terms of  $j_{p\,\mu}^{\alpha}$.  Clearly it would be desirable to devise a new variable that automatically guarantees the gauge-invariance of an approximate $E_{xc}$.  From the work of the previous section we clearly see that the covariant vorticity fulfills our needs.   
We thus write $E_{xc}[j_{p\,\mu}^{\alpha}]=\bar{E}_{xc}[n,{\bf s},\Nubold]$ as anticipated in Eq.~(\ref{ExcJ}).  Additional requirements of rotational invariance forces $\bar{E}_{xc}$ to actually depend on scalar combinations of ${\cal V}_{i}^{\alpha}$ as we will see in an explicit example below.

Given an approximate functional of this form the single particle Kohn-Sham (KS) equation can be written as~\cite{ref:bencheikh}
\begin{equation}\label{eq:KS}
\left[ \frac{1}{2}\sum_i\left(-i \partial_i +{\cal A}_{{\rm KS},i}^{\alpha}\tau^{\alpha}\right)^2+{\cal A}_{{\rm KS},0}^{\alpha}\tau^{\alpha}\right]\Phi_{\lambda}=\epsilon_\lambda \Phi_{\lambda}\,,
\end{equation}
Here $\Phi_{\lambda}({\bf r})= (\phi_{\lambda,\uparrow}({\bf r}), \phi_{\lambda,\downarrow}({\bf r}))^T$ is a two-component spinor with $\phi_{\lambda,\sigma}({\bf r})$ being a single-particle wave function, and the KS potentials are defined as
\begin{align}\label{eq:A_ks}
{\cal A}^0_{{\rm KS},0}&={\cal A}^0_0+v_{\rm H}+{\cal A}^0_{xc,0}
-\frac{1}{2}{\cal A}^\alpha_{xc,i}\left(2{\cal A}^\alpha_i+{\cal A}^\alpha_{xc,i}\right)\,,\nonumber\\
{\cal A}^a_{{\rm KS},0}&={\cal A}^a_0+{\cal A}^a_{xc,0}-{\cal A}^a_i{\cal A}^0_{xc,i}
-{\cal A}^a_{xc,i}\left({\cal A}^0_i +{\cal A}^0_{xc,i}\right)\,,\nonumber\\
{\cal A}^{\alpha}_{{\rm KS},i}&={\cal A}^{\alpha}_{i}+{\cal A}^{\alpha}_{xc,i}\,,
\end{align}
in order to reproduce same paramagnetic spin-current-densities as the interacting system. Here the exchange-crrelation potentials are
\begin{equation}\label{eq:a_xc}
{\cal A}_{xc,\mu}^{\alpha}=\frac{\delta \bar E_{xc}[n,{\bf s},\Nubold]}{\delta j_{p\,\mu}^{\alpha}}\,,
\end{equation}
and the Hartree potential is
\begin{equation}\label{eq:v_h}
v_{\rm H}=\frac{\delta E_{\rm H}[n]}{\delta n}\,.
\end{equation}
The paramagnetic spin-current densities can be readily obtained from KS orbitals:
\begin{align}\label{eq:j_ks}
j_{{\rm p}\,0}^{\alpha}({\bf r})&=s^{\alpha}({\bf r})=\sum_{\lambda}\Phi_{\lambda}^{\dagger}({\bf r})\tau^{\alpha}\Phi_{\lambda}({\bf r})\,, \nonumber\\
j_{{\rm p}\,i}^{\alpha}({\bf r})&=\frac{-i}{2}\sum_{\lambda}\left [ \Phi^{\dagger}_{\lambda}({\bf r})\tau^{\alpha} \partial_{i}\Phi_{\lambda}({\bf r}) -\partial_i \Phi_{\lambda}^{\dagger}({\bf r}) \tau^{\alpha} \Phi_{\lambda}({\bf r})\right] \,,
\end{align}
where sums run over occupied orbitals.

\section{Local density approximation}\label{lda}
While the theory of the previous section is formally exact, it is not yet obvious that a local density approximation  can be naturally formulated in terms of $\Nubold$.  To show that this is indeed the case we follow the approach of Ref. \onlinecite{ref:vignale90}, i.e. we start from the form of exchange-correlation energy and potentials for an electron gas of uniform density subjected to {\it small} and {\it uniform} SU(2) potentials ${\cal A}_i^a$. This system can be considered as a reference for LDA calculations.
The leading order correction to the ground-state energy for weak ${\cal A}_i^a$ is 
\cite{ref:tokatly08} 
\ber\label{eq::e_f2}
E_{\rm GS}&=&E_{\rm GS}^0+\frac{\lambda}{8}\int \mathrm{d}{\bf r}\,{\rm Tr}\left\{\bar{\cal B}_{i}\bar{\cal B}_{i}\right\}\,,\nn\\
&=&E_{\rm GS}^0+\frac{\lambda}{8}\int \mathrm{d}{\bf r}\,{\cal B}^{a}_{i}{\cal B}^{a}_{i}\,.
\eer
where $\lambda$ is an interaction-dependent constant and $E_{\rm GS}^0$ is the $\Abold$-independent part of the ground state energy.  From Eq.~(\ref{CovariantCurl}) we see that the uniform field is related to the uniform SU(2) vector potentials as follows: 
\be\label{BUniform}
{\cal B}_i^\alpha =- \epsilon_{ijk}\epsilon_{\alpha b c}{\cal A}_j^b{\cal A}_k^c\,.
\ee
Notice that at this level of approximation, there is no difference between using the covariant field ${\Fbold}$ or the invariant ones $\Bbold$: because all the fields are evaluated at the same point in space the link operators cancel in the trace.  The situation would be different if we had a nonlocal functional, but even in that case the difference between $\Bbold$ and $\Fbold$ can be ignored as long as one works to second order in the field strength.  Notice, however,  that, in writing Eq.~(\ref{eq::e_f2})  we have assumed that the energy is an analytic function of  ${\Fbold}$: the latter appears quadratically as required by rotational invariance. 

The form (\ref{eq::e_f2}) of the energy functional, combined the above relation between $\Fbold$ and $\Abold$, and with Eq.~(\ref{SpinCurrentDefinition}),  implies the existence of a uniform equilibrium spin current, which cubically depends on the potentials\cite{ref:tokatly08}:  
\be
j_i^a=\lambda\left({\cal A}_i^a {\cal A}_j^b {\cal A}_j^b- {\cal A}_j^a {\cal A}_i^b {\cal A}_j^b\right)\,.
\ee 
Hence the paramagnetic spin current is given by the expression
\begin{equation}
j_{p\,i}^a=\lambda\left({\cal A}_i^a {\cal A}_j^b {\cal A}_j^b- {\cal A}_j^a {\cal A}_i^b {\cal A}_j^b\right)
-{\cal N}^{a\beta} {\cal A}^{\beta}_i\,,
\end{equation}
which, to leading order in the potentials (i.e., to first order in SO interaction), reduces to the following simple form
\be
j_{p\,i}^a \approx-{\cal N}^{a\beta} {\cal A}^{\beta}_i\,.
\ee
It is clear from this expression that the field strength coincides (at this level of approximation)  with the covariant vorticity.  Thus, the ground-state energy  functional for weak fields  is obtained simply by replacing ${\cal B}^{a}_{i} \to {\cal V}^{a}_{i}$ in Eq.~(\ref{eq::e_f2}) and the corresponding xc energy functional is
\ber\label{eq:exc_nu}
E_{xc}&=&E^0_{xc}+\frac{\lambda_{xc}}{8} \int \mathrm{d}{\bf r}\, {\rm Tr}\{\bar{\cal V}_{i}(\rv)\bar{\cal V}_{i}(\rv)\}\nn\\
&=&E^0_{xc}+\frac{\lambda_{xc}}{8} \int \mathrm{d}{\bf r}\,{\cal V}^{a}_{i}(\rv){\cal V}^{a}_{i}(\rv)\,,
\eer 
where  $E^0_{xc}$ is the xc energy of the system with zero vorticity, and ${\cal V}^{a}_{i}$ is given, in the limit of slowly varying densities and currents,  by the last term of Eq.~(\ref{CovariantVorticity2}), i.e.,   ${\cal V}_i^{a}\simeq -\epsilon_{ijk}\epsilon_{a b c}v_j^b v_k^c$
(all indices spatial),  and $\lambda_{xc}=\lambda-\lambda_0$ is the difference between the fully interacting $\lambda$ and the noninteracting one.  From Eq.~(\ref{eq:exc_nu})  the xc potential can be easily calculated  according to Eq.~(\ref{eq:a_xc}).  For example, in the absence of spin polarization we get
\begin{equation}\label{AxcLDA}
{\cal A}_{xc,i}^{a}=  \frac{\lambda_{xc}}{n^4}\left(j_{p\,i}^a j_{p\,j}^b j_{p\,j}^b-j_{p\,j}^a j_{p\,i}^b j_{p\,j}^b\right)\,.\\
\end{equation}
While Eq.~(\ref{eq:exc_nu})  is of the same form as the first CDFT functional proposed for weak fields in Refs.~\onlinecite{ref:vignale87} and \onlinecite{ref:vignale88}, we point out that the quadratic functional dependence of the covariant vorticity on paramagnetic current produces a completely different and new dependence of ${\cal A}_{xc}$ on $j_{p}$~\cite{foot:ordinaryxc}.

\section{Calculation of $\lambda_{xc}$ in a two-dimensional electron gas}\label{rashba}
\begin{figure}
\begin{center}
\includegraphics[width=1.0\linewidth]{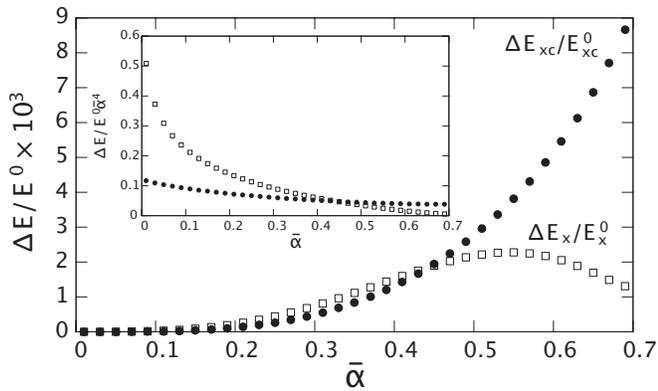}
\caption{Excess exchange-correlation energy $\Delta E_{xc}$ (filled circles) and excess exchange energy $\Delta E_x$ (empty rectangles) of a two-dimensional electron gas at Wigner-Seitz radius $r_s=1$ vs Rashba spin-orbit coupling constant ${\bar \alpha}=\alpha/v_{\rm F}$  ($v_{\rm F}=\sqrt{2}/r_s$ is the Fermi velocity) in units of $E^0_{xc} = -0.8$ and $E^0_x =-0.6$ (in the atomic units) respectively.  
\emph{Inset:} same as in the main figure but in units of $E^0_{xc} {\bar \alpha}^4$ and $E^0_{x} {\bar \alpha}^4$, respectively for the exchange-correlation and exchange energies. 
\label{fig:one}}
\end{center}
\end{figure}

\begin{figure}
\begin{center}

\includegraphics[width=1.0\linewidth]{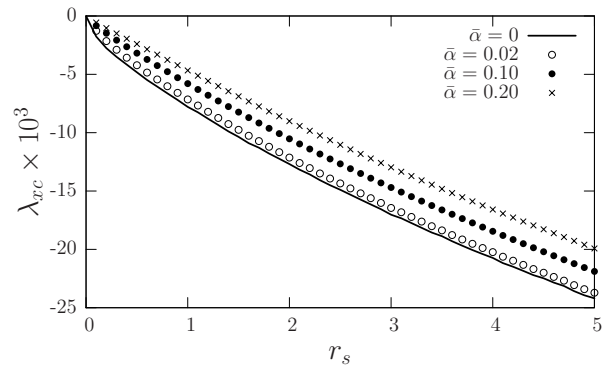}
\caption{$\lambda_{xc}$ of a two-dimensional electron gas with Rashba spin-orbit coupling vs Wigner-Seitz radius $r_s$, for varying Rashba coupling constant ${\bar \alpha}=\alpha/v_{\rm F}$ ($v_{\rm F}=\sqrt{2}/r_s$). The limiting behavior at ${\bar \alpha}=0$ (solid line) is obtained by extrapolation.
\label{fig:two}}
\end{center}
\end{figure}

Let us finally outline the calculation of $\lambda_{xc}$ for a uniform two-dimensional electron gas with linear spin-orbit coupling (Rashba model~\cite{ref:bychkov84}).  The model in question~\cite{foot:rashba} is characterized by a gauge potential with components 
\be
{\cal A}_2^1=-{\cal A}_1^2= \alpha\,,
\ee
where $\alpha$ is the Rashba coupling constant. This implies that
\be
{\cal B}_{3}^{3}=-2 \alpha^2\,,
\ee
is the only non-zero component of $\Fbold$, and from our previous discussion we expect  $\Delta E_{xc} \equiv E_{xc}-E^0_{xc} \propto \alpha^4$.  
Some  details of the calculations are provided in Appendix~\ref{app:rashba}.  An interesting observation is that the exchange energy alone, while clearly gauge invariant,  is a {\em non-analytic} function of $\alpha$, being  $\propto\alpha^4\ln{\alpha}$ at small $\alpha$~\cite{ref:chesi_thesis}. So it is essential to include the correlation energy in order to restore the expected analyticity of the energy functional. This we have done with the help of the \emph{random phase approximation} (RPA)~\cite{ref:vignale_book}.  In Fig.~\ref{fig:one} we plot the ratio $\Delta E_{xc}/E^0_{xc}$ vs $\alpha$   as well as the corresponding results for the exchange only. 
It is immediately evident that $\Delta E_{xc}$ scales as $\alpha^4$ but $\Delta E_x$ does not.  In Fig.~\ref{fig:two} we plot the constant $\lambda_{xc}$ as a function of Wigner-Seitz radius $r_s$.   This quantity will be the necessary input for LDA calculations within our spin-current-density functional formalism.


\section{Discussion}\label{cncl}
The results of this paper are primarily formal.  The introduction of the ``invariant vorticity" allowed us to provide a formal solution to a long-standing problem in spin-CDFT, namely, the problem of constructing manifestly SU(2)-invariant functionals of the paramagnetic current densities.  An interesting feature of the present construction is that the use of the paramagnetic spin current as basic variables is mandatory: without them we could not define the link operators $\hat L_v(\rv)$ (see Eq.~(\ref{defvlink})), which are vital to the definition of the invariant vorticity.  This is at odds with spinless CDFT,  in which the use of the paramagnetic current  density as a basic variable is, in a sense, optional.  That is, while the use of the paramagnetic current  is strongly indicated for practical reasons (e.g., for constructing a local density approximation),    one might nevertheless chose with work with the physical current, provided one is willing to accept a much higher degree of nonlocality in the functionals.  The situation is quite different here.  It is precisely the nonlocality of the functionals that forces us to introduce the link operator, and hence the paramagnetic current.  However, this complication does not show up within the local density approximation.

Even more striking is the fact that the paramagnetic spin current must necessarily enter the construction of a nonlocal SU(2)-invariant functional of the {\it spin density}.  In other words, such a functional must be constructed in terms of the invariant spin density $\hat{\bar s}(\rv)$ of Eq.~(\ref{InvariantS}), rather than the ordinary spin density.
The physical reason for this is that SU(2) symmetry cannot be maintained without the introduction of SU(2) vector potentials, which in turn modify the paramagnetic spin current densities.  A truly SU(2)-invariant nonlocal spin DFT must necessarily be a spin-current DFT.  Even if we stop at the ``semi-local" level, e.g. by doing a gradient expansion in the spin density, SU(2) invariance requires that the gradients be replaced by covariant gradients, which are most naturally defined in terms of the connection $\hat v_i$.  Another possibility is to resort to functionals of spin-orbitals, which can be SU(2) invariant without explicitly involving spin currents.

In spite of the formal progress, there is a major weakness in our proposal.   The link operators depend, for their definition, on an arbitrary choice of linking paths (see Fig.~\ref{Fig0}).   While it is evident that physical results cannot depend on the choice of the paths, we cannot exclude the possibility that the form of an approximate functional might depend on this choice, even though the functional itself is SU(2) invariant for a given choice.  This question clearly deserves further consideration.  

In conclusion, we hope that our results will stimulate further work aimed at constructing  novel exchange-correlation functionals for spin-orbit coupled systems.  

\section{Acknowledgments}
We would like to thank R. Asgari, S. Chesi and M.M. Sheikh-Jabbari for useful discussions. 
This work has been supported by NSF under Grant No. DMR-0705460.  GV gratefully acknowledges support from the IKERBASQUE foundation and the kind hospitality of the ETSF in San Sebastian, Spain,  and the hospitality of the ISSP  of the University of Tokyo, where part of this work was done.  SHA appreciates support and hospitality of the IPM during the final stages of this work.

\appendix
\section {Details for the calculation of $\lambda_{xc}$}\label{app:rashba}
Exchange-correlation energy is obtainable by means of the fluctuation-dissipation theorem.
The excess exchange energy per particle, due to spin-orbit interaction reads~\cite{ref:vignale_book}
\begin{equation}\label{eq:ex}
\Delta E_{x}=\frac{-S}{\left(2\pi\right)^2 }\int_0^{\infty}{\mathrm d}q\,q\, v_q\int_0^{\infty}{\mathrm d}\omega\left[\Pi_{\alpha}(q,i \omega)-\Pi_{0}(q,i \omega)\right]\,,
\end{equation}
and the excess exchange-correlation energy per particle, within the random phase approximation is given by
\begin{equation}\label{eq:rpa_xc}
\Delta E_{xc}^{\rm RPA}=\frac{S}{\left(2\pi\right)^2 }\int_0^{\infty}{\mathrm d}q\,q\int_0^{\infty}{\mathrm d}\omega\ln{\left[\frac{1-v_q\Pi_{\alpha}(q,i \omega)}{1-v_q\Pi_0(q,i \omega)}\right]}\,.
\end{equation}
In Eqs.~(\ref{eq:ex}) and~(\ref{eq:rpa_xc}), $S$ is the area of system, $\Pi_0(q,i \omega)$ is the noninteracting density-density response function of a two-dimensional electron gas (2DEG) without spin-orbit coupling (\emph{i.e.} Stern function\cite{ref:vignale_book}), and $\Pi_{\alpha}(q,i \omega)$, the noninteracting density-density response function of a 2DEG with Rashba spin-orbit coupling is
\begin{equation}\label{rashba_stern}
\Pi_{\alpha}({\bf q},i\omega)=\frac{1}{S}\sum_{{\bf k},\mu,\nu}{\frac{n_{{\bf k},\mu}-n_{{\bf k}+{\bf q},\nu}}{i\omega+\varepsilon_{{\bf k},\mu}-\varepsilon_{{\bf k}+{\bf q},\nu}}}{\cal F}^{\mu\nu}_{{\bf k},{\bf k}+{\bf q}}\,,
\end{equation}
where $n_{{\bf k},\mu}$ is the Fermi distribution function and $\varepsilon_{{\bf k},\mu}=k/2 -\mu \alpha k$. Moreover 
\begin{equation}
{\cal F}^{\mu\nu}_{{\bf k},{\bf k}+{\bf q}}=\frac{1+\mu \nu \cos{\theta_{{\bf k},{\bf k}+{\bf q}}}}{2}\, ,
\end{equation}
is the form factor, with $\theta_{{\bf k},{\bf k}+{\bf q}}$ being the angle between ${\bf k}$ and ${\bf k}+{\bf q}$.
After changing sum over $\kv$ in Eq~(\ref{rashba_stern}) into integral and performing sum over $\nu$, we find
\begin{widetext}
\begin{equation}
{\Pi}_{\alpha}(x,i u)=-\frac{1}{(2\pi)^2}\sum_{\mu}\Re e \left[\int_0^{\beta_\mu}d y\,y\int_0^{2\pi}d\phi\frac{d_{\mu}\cos{\phi}+e_{\mu}}{a\cos^2{\phi}+b_{\mu}\cos{\phi}+c_{\mu}}\right]\,.
\end{equation}
\end{widetext}
Here $y=k/k_{\rm F}$, $x=q/(2 k_{\rm F})$, $u= \omega/(k_{\rm F} q)$, $\beta_\mu=k_{{\rm F}\mu}/k_{\rm F}=\sqrt{1-{\bar \alpha}^2}+\mu{\bar \alpha}$ and ${\bar \alpha}=\alpha/k_{\rm F}$, with $k_{\rm F}=\sqrt{2\pi n}$. Here we have also introduced the following notations:
\begin{align}
a&=x y^2\,, \nonumber\\
b_{\mu}&=y\left[2 x \left(x- i u\right)+{\bar \alpha}\left(\mu y-{\bar \alpha}\right)\right] \,, \nonumber\\
c_{\mu}&=x\left[\left(x- i u\right)^2-{\bar \alpha}^2\right]+\mu{\bar \alpha}y \left(x- i u\right)\,, \nonumber\\
d_{\mu}&=y+\mu{\bar \alpha}\,, \nonumber\\
e_{\mu}&=x- i u+\mu{\bar \alpha}y/x\,.
\end{align}
Integration over $\phi$ can be performed analytically and the result reads
\begin{widetext}
\begin{equation}\label{eq:chai_alpha}
{\Pi}_{\alpha}(x,i u)=\frac{1}{2\pi}\sum_{\mu,\lambda}\int_0^{\beta_\mu}d y\,\Re e \left[\frac{v_{\mu\lambda}+v^*_{\mu\lambda}}{\left|v_{\mu\lambda}+v^*_{\mu\lambda}\right|}\frac{\lambda y \left(d_{\mu}v_{\mu\lambda}+e_{\mu}\right)}{a\left(v_{\mu+}-v_{\mu-}\right)\sqrt{v_{\mu\lambda}^2-1}}
\right]\,,
\end{equation}
\end{widetext}
with $v_{\mu\pm}=\left(-b_{\mu}\pm\sqrt{b_{\mu}^2-4 a c_{\mu}}\right)/(2 a)$.

Now using Eq.~(\ref{eq:chai_alpha}) in Eqs.~(\ref{eq:ex}) and~(\ref{eq:rpa_xc}), and performing the three remaining integrals over $y$, $\omega$ and $q$ numerically, we find spin-orbit correction to the exchange and exchange-correlation energies of a 2DEG.


\begin{thebibliography}{99}
\bibitem{ref:vignale_book}
	G.F. Giuliani and G. Vignale, {\it Quantum Theory of the Electron Liquid}  (Cambridge University Press, Cambridge, 2005).
\bibitem{ref:vignale87}
	G. Vignale and M. Rasolt, Phys. Rev. Lett. {\bf 59}, 2360 (1987).
\bibitem{ref:vignale88}
	G. Vignale and M. Rasolt, Phys. Rev. B {\bf 37}, 10685 (1988).
\bibitem{ref:vignale90}
	G. Vignale, M. Rasolt, and D. J. W. Geldart, Adv. Quantum Chem. {\bf 21}, 235 (1990).
\bibitem{ref:rajagopal73}
	A.K. Rajagopal and J. Callaway, Phys. Rev. B {\bf 7}, 1912 (1973).
\bibitem{ref:eschrig85}
	H. Eschrig, G. Seifert, and P. Ziesche, Solid State Commun. {\bf 56}, 777 (1985).
\bibitem{ref:bencheikh}
	K. Bencheikh, J. Phys. A: Math. Gen.  {\bf 36}, 11929 (2003).
\bibitem{ref:rohra07}
	S. Rohra, E. Engel, and A. G\"orling, arXiv:cond-mat/0608505.
\bibitem{ref:tatara}
	G. Tatara, H. Kohno, and J. Shibata, Phys. Rep. {\bf 468}, 213 (2008).
\bibitem{ref:castroneto}
	A.H. Castro Neto, F. Guinea, N.M. Peres, K.S. Novoselov, and A.K. Geim,
	Rev. Mod. Phys. {\bf 81}, 109 (2009).
\bibitem{ref:rohra06}
	S. Rohra and A. G\"orling, Phys. Rev. Lett. {\bf 97}, 013005 (2006).
\bibitem{ref:helbig08}
	N. Helbig, S. Kurth, S. Pittalis, E. R\"as\"anen, E.K.U. Gross,
	Phys. Rev. B {\bf 77}, 245106 (2008).
\bibitem{ref:tokatly08}
	I.V. Tokatly, Phys. Rev. Lett. {\bf 101}, 106601 (2008).
\bibitem{ref:frohlich93}
	J. Fr\"ohlich and U.M. Studer, Rev. Mod. Phys. {\bf 65}, 733 (1993).
\bibitem{foot:index}
	Here Greek indices go from $0$ to $3$,  while Latin indices go from $1$ to $3$.
	Moreover we are distinguishing three and four vectors with bold characters and overhead arrows, respectively.
\bibitem{ref:bohm}
	A. Bohm, A. Mostafazadeh, H. Koizumi, Q. Niu, and J. Zwanziger,
	 {\it The Geometric Phase in Quantum Systems: Foundations, Mathematical Concepts, and Applications in Molecular and Condensed Matter Physics
} (Springer-Verlag, Berlin, 2003).
\bibitem{foot:fieldstrength} 
The field strength is usually defined as a tensor:
${\cal F}_{\mu\nu}^{\alpha}\tau^{\alpha}={\cal D}_{\mu} {\cal A}_{\nu}^{\alpha}\tau^{\alpha}-{\cal D}_{\nu} {\cal A}_{\mu}^{\alpha}\tau^{\alpha}$.  For our purpose it is simpler to work with the dual vector field ${\cal B}_i^{\alpha}$.  Notice that  ${\cal B}_i^{\alpha}{\cal B}_i^{\alpha}={\cal F}_{ij}^{\alpha}{\cal F}_{ij}^{\alpha}$.
\bibitem{ref:itzykson} 
	C. Itzykson and J.-B. Zuber {\it Quantum Field Theory}, (McGrow-Hill, 1980).
\bibitem{ref:broda2001}
	B. Broda, {\it Modern Nonlinear Optics}, Vol.~2, edited by M.~W.~Evans (Wiley, New York, 2001), p.429; arXiv:math-ph/0012035.
\bibitem{foot:symmetry}
Notice that rotational invariance is an exact property of the energy functional, which continues to hold even if rotational symmetry is broken in the ground state.
 \bibitem{foot:ordinaryxc}
Notice Eq.~(\ref{AxcLDA}) defines only the spin part of the xc vector potential, ${\cal A}_{xc,i}^{a}$.  The ordinary charge component ${\cal A}_{xc,i}^{0}$ is still given by the VR weak field approximation.
\bibitem{ref:bychkov84}
		Y.A. Bychkov and E.I. Rashba, J. Phys. C {\bf 17}, 6039 (1984).
\bibitem{foot:rashba}
 We have added a $\alpha^2/2$ term to the single particle part of the original Rashba Hamiltonian to make it gauge invariant. Evidently this does not affect the exchange-correlation energy.  
\bibitem{ref:chesi_thesis}
	S. Chesi, {\it Effects of Structural Spin-Orbit Coupling in Two Dimensional Electron and Hole Liquids} (Ph.D. Thesis, Purdue University, 2007).
\end{thebibliography}
\end{document}